\documentclass[12pt,preprint]{aastex}

\begin{document}

\def\deg{^{\circ}}

\title{PSR J1756$-$2251: a new relativistic double neutron
star system} 
\author{ A.~J.Faulkner\altaffilmark{1}, M. Kramer\altaffilmark{1},
A.~G. Lyne\altaffilmark{1}, R.~N. Manchester\altaffilmark{2}, 
M.~A. McLaughlin\altaffilmark{1},\\ 
I.~H. Stairs\altaffilmark{3}, G. Hobbs\altaffilmark{2}, 
A. Possenti\altaffilmark{4}, D.~R. Lorimer\altaffilmark{1}, \\
N. D'Amico\altaffilmark{4,5}, F. Camilo\altaffilmark{6} 
\& M. Burgay\altaffilmark{4}} 

\altaffiltext{1}{University of Manchester, Jodrell Bank Observatory, Macclesfield,
Cheshire, SK11~9DL, UK}
\altaffiltext{2}{Australia Telescope National Facility, CSIRO, 
P.O.~Box~76, Epping NSW~1710, Australia} 
\altaffiltext{3}{Department of Physics \& Astronomy, University of 
British Columbia, 6224 Agricultural Road, Vancouver, B.C. V6T 1Z1, Canada } 
\altaffiltext{4}{INAF - Osservatorio Astronomico di Cagliari, Loc. 
Poggio dei Pini, Strada 54, 09012, Capoterra (CA), Italy } 
\altaffiltext{5}{Universita' degli Studi di Cagliari, Dipartimento di
Fisica, SP Monserrato-Sestu km 0,7, 90042, Monserrato (CA), Italy} 
\altaffiltext{6}{Columbia Astrophysics Laboratory, Columbia
University, 550 West 120th Street, New York, NY~10027, USA}

\begin{abstract} 

We report the discovery during the Parkes Multibeam Pulsar Survey of
PSR J1756$-$2251, a 28.5~ms pulsar in a relativistic binary
system. Subsequent timing observations showed the pulsar to have an
orbital period of 7.67~hrs and an eccentricity of 0.18. They also
revealed a significant advance of periastron, $2.585 \pm 0.002$
deg. yr$^{-1}$. Assuming this is entirely due to general relativity
implies a total system mass (pulsar plus companion) of
2.574$\pm$0.003~M$_{\odot}$. This mass and the significant orbital
eccentricity suggest that this is a double neutron star
system. Measurement of the gravitational redshift, $\gamma$, and an
evaluation of the Shapiro delay shape, $s$, indicate a low companion
mass of $<$1.25~$M_{\odot}$. The expected coalescence time due to
emission of gravitational waves is only $\sim$1.7~Gyr substantially
less than a Hubble time. We note an apparent correlation between spin
period and eccentricity for normally evolving double neutron star
systems.

\end{abstract}


\clearpage

\section{Introduction}

Relativistic double neutron star (DNS) binary systems in tight orbits
are valuable physical laboratories, since their rapid evolution allows
stringent tests of gravitational theories in strong-field conditions
(e.g., Taylor \& Weisberg 1989\nocite{tw89}). They make a significant
contribution to estimated event rates for gravitational wave detectors
(e.g. Kim, Kalogera \& Lorimer 2002\nocite{kkl03}). DNS binaries are
rare since, even if the system survives the first supernova explosion,
the progenitor system will typically disrupt with the second supernova
explosion. Also, they are observationally selected against because of
the large orbital modulation of the pulsar period. In this {\it
Letter} we report the discovery of PSR J1756$-$2251, a 28.5~ms pulsar,
and present strong evidence which suggests that it is the 8th DNS
binary system known and the 5th which will coalesce in less than a
Hubble time. In \S 2 we describe the observations, \S 3 evaluates the
nature of the companion, while \S 4 discusses the search for a
possible pulsar companion. Finally, \S 5 considers some of the
implications of the discovery.

\section{Observations}
\label{s:observations}

PSR J1756$-$2251 was discovered in the Parkes Multibeam Pulsar Survey
(PMPS; Manchester et al. 2001\nocite{mlc+01}). The PMPS is the most
successful survey for pulsars, with more than 700 pulsars discovered
so far. The survey used a sensitive 13-beam receiver at the Parkes
radio telescope to cover the Galactic plane $(|b|<5^{\circ}$,
$260^{\circ}<l<50^{\circ})$ with 35-min integrations at a centre
frequency of 1374~MHz, with 96 frequency channels covering 288~MHz
bandwidth, using 250~$\mu$s sampling.

An initial search using standard Fourier techniques, as described in
Manchester et al. (2001)\nocite{mlc+01}, failed to detect PSR
J1756$-$2251. The data were subsequently re-analysed using an
efficient acceleration code (Faulkner et al. 2004\nocite{fsk+04}) and
PSR J1756$-$2251 was detected in three beams, with the best
signal-to-noise ratio of 19. The system was confirmed in June 2003 by
re-observation at Parkes, using the `gridding' technique described in
Morris et al. (2002\nocite{mhl+02}).

\begin{figure}
\includegraphics[scale=0.7]{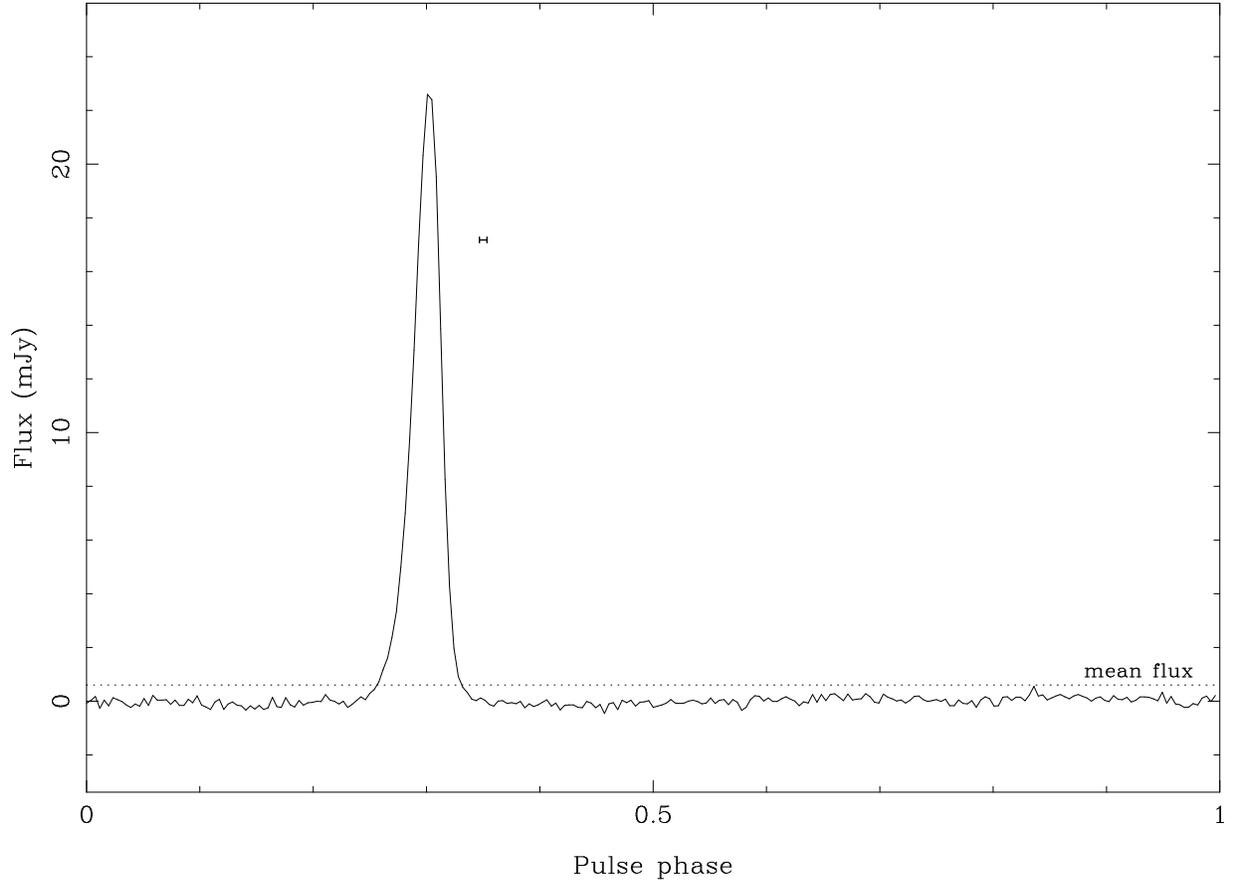}
\caption{Average pulse profile of PSR J1756$-$2251 at 1390~MHz,
obtained by integrating 27.9 hours of observation at Parkes. The small
horizontal bar to the right of the pulse indicates the resolution of
the profile, including the effects of interstellar dispersion}
\label{f:profile}
\end{figure}

PSR J1756$-$2251 has subsequently been observed regularly from Parkes
at 1374~MHz using 288~MHz bandwidth (BW) and 1390~MHz using 256~MHz BW
and 80 or 125~$\mu$s sampling time ($t_{\rm s}$), plus occasional
observations at 685~MHz (64~MHz BW, $t_{\rm s}$: 80~$\mu$s) and
3030~MHz (768~MHz BW, $t_{\rm s}$: 80~$\mu$s); at Jodrell Bank
Observatory at 1396~MHz (64~MHz BW, $t_{\rm s}$: 371~$\mu$s) and
610~MHz (4~MHz BW, $t_{\rm s}$: 555~$\mu$s) and the Green Bank
Telescope (GBT) at 1400~MHz (96~MHz BW, $t_{\rm s}$: 72~$\mu$s) using
the Berkley Caltech Pulsar Machine (Backer et
al. 1997\nocite{bdz+97}). Full orbit observations of $\sim$8~hrs have
been made at Parkes and the GBT. Figure \ref{f:profile} shows the
integrated pulse profile of PSR J1756$-$2251, at 1390~MHz with 512~kHz
frequency channels and 80~$\mu$s sampling; it is sharp but
featureless, with a pulse width of 2.7\% of the spin period. The 1998
observations, with 3~MHz frequency channels, have higher dispersion
smearing, however, the profile still appears similar. The profile at
3030~MHz is also similar while there is broadening at 685~MHz. Given
that profile widths of recycled pulsars hardly evolve with frequency
(Kramer et al. 1999\nocite{kll+99}), this broadening is likely to be
due to scattering. There is no evidence of any emission at phase
180$^{\circ}$ from the pulse.

For regular coverage over the orbit, pulse time of arrivals (TOAs)
were made using 10~min sub-integrations from the Parkes
observations. Three survey observations, made 5 years before the
confirmation, were included. The bulk of the observations, including
1189 of the 1382 timing points, were made at Jodrell Bank, each using
5~min of integration each. The timing analysis used the \textsc{TEMPO}
program\footnote{See http://www.atnf.csiro.au/research/pulsar/tempo/}
and gave the timing parameters listed in Table \ref{t:parameters}. PSR
J1756$-$2251 is very close to the ecliptic plane, which has made the
determination of its declination using timing measurements relatively
imprecise.

\section{Nature of the Companion}
\label{s:nat_comp}

The masses of the pulsar and its companion cannot be measured
independently in purely Keplerian orbits. However, the `mass
function', $f_{\rm mass}={(m_2\ {\rm sin}\,i)^3}/{(m_1+m_2)^2}$, can be
found from the orbital period $P_{\rm b}$ and projected semi-major
axis, $x=a_1\ {\rm sin}\,i$, where $i$ is the orbital inclination. This
relates the masses of the pulsar, $m_1$, and the companion, $m_2$,
both in solar masses. Since we fit for $0^{\circ}\leq i \leq
90^{\circ}$, clearly cos$\,i$ must be $>$0.

We have measured the precession of the longitude of periastron,
$\dot\omega$, to be $2.585 \pm 0.002$~deg~yr$^{-1}$. This could be due
to a combination of tidal effects caused by a non-compact deformable
companion (Smarr \& Blandford 1976\nocite{sb76}), the effects of
relativistic gravitational interaction (Taylor \& Weisberg
1989\nocite{tw89}) or spin orbit coupling (e.g. Masters \& Roberts
1975)\nocite{mr75}. Although it would be possible for a hydrogen
main-sequence star to fit in the orbit, it would induce an
$\dot\omega$ of $\sim$1000 times the observed value (see Masters \&
Roberts 1975)\nocite{mr75}. Alternatively, the companion could be a
helium main-sequence star. Using the relationship between
$\dot\omega$, $m_{\rm 1}$ and $m_{\rm 2}$ due to tidal effects derived
in Roberts, Masters \& Arnett (1976)\nocite{rma76}, we find that the
required pulsar mass must be $<0.6$~$M_{\odot}$, which is very
unlikely. Furthermore, there is no evidence of occultations of the
pulsar, which is consistent with the companion's identification as a
compact object. Hence, the companion must be a compact object; either
a neutron star (NS) or white dwarf (WD). A fast rotating WD could
affect $\dot\omega$ by spin orbit coupling. Following Wex (1998)
\nocite{wex98}, we estimated the size of the contribution from a
derived upper limit on $\dot{x}/x$.  While we do not know all relevant
angles in the system's geometry, we conclude that this effect is
insignificant, unless we observed the system under very specific
angles and/or at particular phases of the orbital precession. While we
consider this as unlikely, future observations will verify this
conclusion.

\begin{figure}
\includegraphics[scale=0.7]{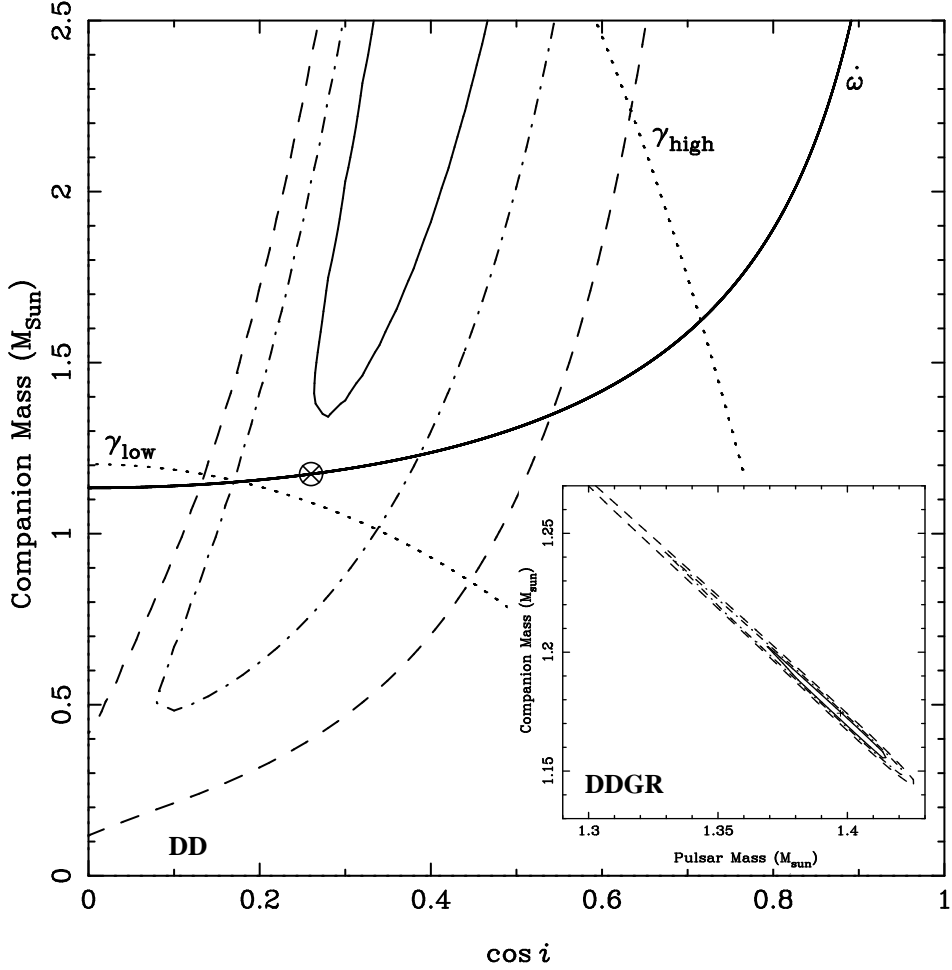} 
\caption{
\label{f:m2cosi}
The observational constraints, using the DD model, on the companion
mass and orbital inclination, $i$, shown as cos$\,i$. The value of
$\dot\omega$ is well constrained and the solid line shows both the
upper and lower limits of $\dot\omega$. The dotted lines show the much
wider limits imposed by $\gamma$. Both $\dot\omega$ and $\gamma$
errors are twice the formal \textsc{TEMPO} errors. The three contours
show the 1-$\sigma$, 2-$\sigma$ and 3-$\sigma$ ranges for a fixed
Shapiro $s$ parameter, and companion mass, with all other parameters
left free. The insert shows a contour plot, also 1-$\sigma$,
2-$\sigma$ and 3-$\sigma$, of companion and pulsar masses using the
DDGR model: see text for details. The best fit companion mass found is
shown as a cross on the main plot.
}
\end{figure}

Taking $\dot\omega$ to be entirely due to the effects of general
relativity, the total system mass can be derived (see Damour \& Taylor
1992)\nocite{dt92}. We obtain a value of $2.574 \pm
0.003$~$M_{\odot}$. Figure \ref{f:m2cosi} shows $\dot\omega$ as a
function of $m_2$ and cos$\,i$; the data were analysed using
the DD model of \textsc{TEMPO} (Damour \& Deruelle 1985,
1986\nocite{dd85}\nocite{dd86}). The total system mass is remarkably
similar to that of the double pulsar system, J0737$-$3039
($2.588$~M$_{\odot}$; Burgay et al. 2003\nocite{bdp+03}).

The system mass, age of the pulsar and the significant eccentricity of
0.18 make a WD companion unlikely, since any accretion from the WD
progenitor onto the NS would have circularised the orbit. If the
companion is a WD then it would be a CO type with a mass of
$\sim$1.2~$M_{\odot}$. At a distance of 2.5~kpc and assuming the
pulsar characteristic age of 443~Myr we expect it to be magnitude
$\lesssim$24 (e.g., Gar\'{c}ia-Berro et
al. 1996\nocite{ghi+96}). While it is probably too faint to be seen in
available optical surveys, we have, reviewed surveys and catalogues at
the Astrophysical Virtual
Observatory\footnote{http://www.euro-vo.org}, which includes objects
with a maximum magnitude of 22.5. No plausible optical counterpart was
found.

Although not yet well constrained, we have measured the gravitational
redshift parameter, $\gamma$, (Damour \& Duerelle 1986) to be
$1.3\pm0.3$~ms. This puts further limits on the masses as shown in
Figure \ref{f:m2cosi}.

Measurements of the Shapiro delay parameters: range $r$ and shape $s$,
can provide estimates of both $m_2$ and $i$ (e.g., Damour \& Taylor
1992)\nocite{dt92}. It has not yet been possible to constrain $r$
well, but constraints on $s$ can be obtained by searching the $\chi^2$
hypersphere, obtained by running \textsc{TEMPO} over a range of
sin$\,i$ and $m_2$ values. This further constraint is plotted as
probability contours in Figure \ref{f:m2cosi}.

To constrain the masses of the pulsar and companion further we can
assume that general relativity is correct and apply the DDGR
(Damour-Deruelle General Relativity) model (Taylor \& Weisberg
1989\nocite{tw89}). We have explored the $\chi^2$ hypersphere of
$m_1+m_2$ and $m_2$ which are the only unknown parameters if the orbit
is completely described by general relativity. A contour plot, of
$m_1$ and $m_2$, is shown as an insert in Figure \ref{f:m2cosi}. This
indicates a light neutron star companion, possibly with a mass similar
to or lighter than that of PSR J0737$-$3039B (Lyne et
al. 2004\nocite{lbk+04}).

The expected value of the gravitational wave damping of the orbit is
below our present detection limit. Assuming the pulsar has a mass
1.35~$M_{\odot}$, general relativity predicts a value for $\dot
P_{\rm b}$ of $-2.2\times 10^{-15}$.

\section{Companion Pulsar Search}
\label{s:comp_srch}

PSR J1756$-$2251 probably followed a typical DNS evolution
(e.g. Phinney \& Kulkarni 1994\nocite{pk94}). Both stars would
initially have had masses $>8 M_{\odot}$. The pulsar was formed in a
supernova explosion of the more massive component. Subsequently, the
companion expanded in its red giant phase and the pulsar accreted
material, thus increasing its rotation rate. The two stars then
spiralled together in a common envelope to give a tight
orbit. Finally, the companion star exploded as a supernova.

The possibility that the companion star is also a pulsar has been
explored. Any second pulsar would be highly accelerated, which would
normally have the effect of reducing search sensitivity due to pulse
smearing. However, we now know the precise orbital parameters and
dispersion measure of PSR J1756$-$2251, we also know the ephemeris for
a companion except for the projected semi-major axis, $a_2\ {\rm
sin}\,i$. This is dependent upon the ratio of the masses of
the two stars:
\begin{equation}
a_2\ {\rm sin}\,i=\frac{m_1}{m_2}a_1\ {\rm sin}\,i.
\end{equation}
We are, therefore, able to make a fully coherent search for a
companion pulsar using a series of values of $m_2$ appropriate to a
neutron star. For each search, a time series in the pulsar rest frame
was constructed by taking each time sample from the closest sample in
the observation time series. Corrected time series from two full
orbits of observations, at 1390~MHz from Parkes, were searched using
the standard approach for a solitary pulsar (e.g. Manchester et
al. 2001\nocite{mlc+01}). PSR J0737-3039B is only visible for short
periods at particular orbital phases (Lyne et al. 2004); consequently,
searches were also conducted throughout the observation with a range
of observation lengths from 10~min to the full orbit, with a limiting
flux density of $\sim$0.045~mJy. The process was repeated over a range
of masses: $0.73m_1<m_2<1.27m_1$. The searches were sensitive to
spin-periods from $~$1~ms up to 10~s.

Unfortunately, no companion pulsar was detected with a luminosity
limit of $\sim$0.3~mJy kpc$^2$, below that of the faintest known
pulsar which has a luminosity of 0.5~mJy kpc$^2$ (Camilo
2003\nocite{cam03}).

\section{Discussion}
\label{s:discussion}

Based on orbital damping due to emission of gravitational waves in
general relativity (Peters \& Mathews 1963\nocite{pm63}), the
coalescence time of PSR J1756$-$2251 is $\sim$1.7~Gyr. This is
substantially less than the age of the Universe and this system is
therefore important in the estimation of coalesence rates of DNS
systems.

General relativity predicts that the merging of two neutron stars will
produce a burst of gravity waves (Misner, Thorne \& Wheeler
1973\nocite{mtw73}) detectable over inter-galactic distances by
ground-based gravity-wave (GW) detectors. The pulsars predicted to
coalesce within 10~Gyr are listed in Table \ref{t:coalescing}.

The rate of mergers observable by GW detectors has been discussed
extensively, see e.g. Phinney (1991)\nocite{phi91} and Kim et
al. (2003)\nocite{kkl03}. Following the discovery of PSR J0737-3039,
predicted NS--NS coalesce rates were substantially increased (Burgay
et al. 2003\nocite{bdp+03}, Kalogera et
al. 2004)\nocite{kkl+04}. However, these calculations use only three
systems, PSR's B1913+16, B1534+12 and J0737$-$3039A; PSR B2127+11C is
usually not included in the calculations since it is in a globular
cluster and probably formed by exchange interaction rather than binary
evolution (Prince et al. 1991\nocite{pakw91}). The predicted rate of
mergers is dominated by PSR J0737$-$3039A due to its proximity, short
time to coalesence and difficulty of discovery due to Doppler
smearing. Because its parameters are similar to those of known pulsars
and does not represent a new population, the addition of PSR
J1756$-$2251 is not expected to make a significant difference to the
predicted merger rate (Kalogera et al.~2004).

\begin{figure}
\includegraphics[scale=0.6]{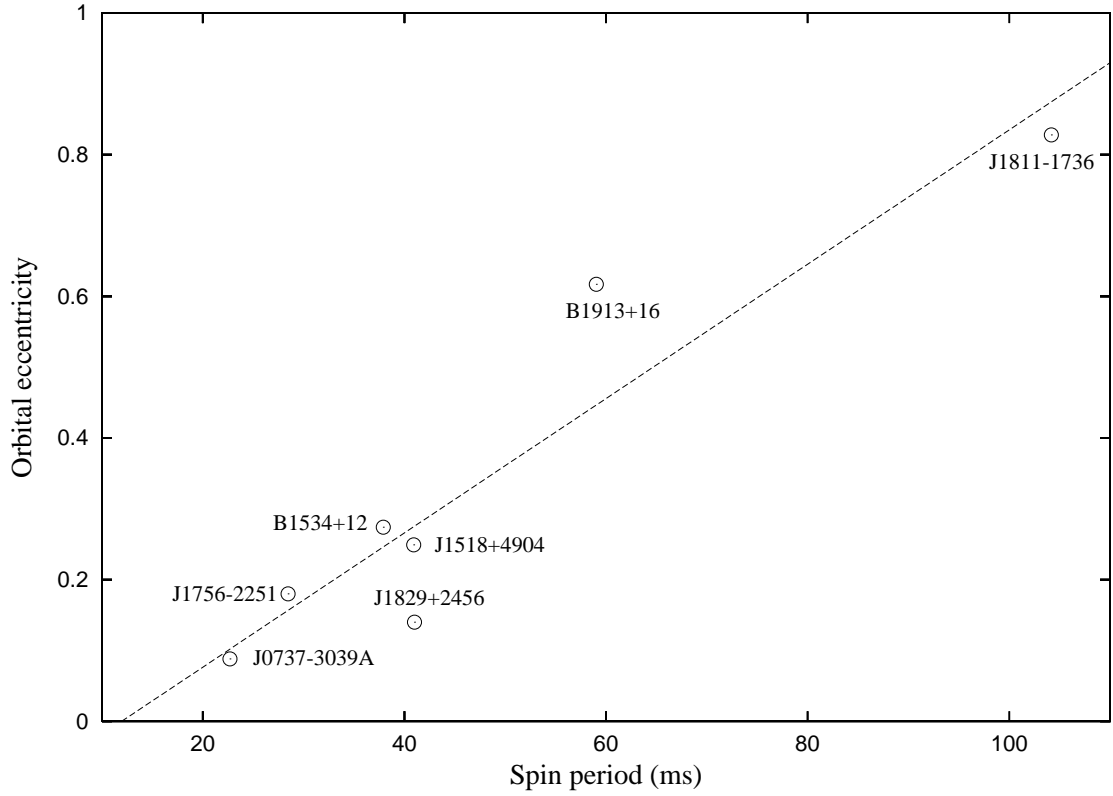} 
\caption{
\label{f:correlation}
Plot showing the strong relationship between eccentricity and spin
period of known DNSs. The dashed line is the best fit for all systems
except PSR B2127+11C (not shown) which is likely to have a non-typical
formation history.
}
\end{figure}

As noted by McLaughlin et al.~(2004)\nocite{mlc+04}, for seven of the
DNS systems known, (excluding PSR B2127+11C, see above), there appears
to be a strong correlation between spin period, $P$, and eccentricity,
$e$, as shown by the dashed line in Figure \ref{f:correlation}. The
Pearson correlation coefficient, $r$, for these seven systems is
0.97. A Monte Carlo simulation in which seven data points are drawn
from a flat distribution in $P$ and $e$ show that such high values of
$r$ occur by chance only 0.1\% of the time. This apparent correlation
shows the current state of the DNS systems. The basic relationship
between $P$ and $e$ was formed early in the system's history, with
some evolution over the actual, but uncertain, age of the
system.

Qualitatively, this relationship may be due to a less massive
progenitor of the companion neutron star evolving through its giant
phase relatively slowly, leaving a longer time to spin-up the pulsar
by accretion, hence a shorter spin period. Prior to the supernova
explosion of the companion star the orbit is expected to have been
circularised due to accretion (e.g Bhattacharya \& van den Heuvel,
1991)\nocite{bv91}. A less massive companion star will eject less mass
from the explosion, hence, using the simplest symmetric mass loss
model which ignores kick velocities, will result in a smaller recoil
(Phinney \& Kulkarni, 1994)\nocite{pk94}. Hence, a less massive
companion star will lead to a less eccentric DNS system as well as to
a shorter spin period. There is a similar correlation between spin
period and companion mass further supporting this mechanism. A more
detailed investigation of the various effects is clearly required.

\section{Acknowledgements} 

 We gratefully acknowledge the technical assistance with hardware and
 software provided by Jodrell Bank Observatory, CSIRO ATNF,
 Osservatorio Astronomico di Cagliari, Swinburne centre for
 Astrophysics and Supercomputing.  The Parkes radio telescope is part
 of the Australia Telescope which is funded by the Commonwealth of
 Australia for operation as a National Facility managed by CSIRO. IHS
 holds an NSERC UFA and is supported by a Discovery Grant. DRL is a
 University Research Fellow funded by the Royal Society.  FC
 acknowledges support from NSF grant AST-02-05853 and a NRAO travel
 grant. NDA, AP and MB received support from the Italian Ministry of
 University and Research (MIUR) under the national program {\it Cofin
 2003}. We thank Thomas Driebe for calculations of white dwarf
 luminosities.

\clearpage

\begin{deluxetable}{ll}
\tabletypesize{\footnotesize}
\tablewidth{0pt} 
\tablecaption{Observed and derived characteristics of PSR J1756$-$2251.}
\tablehead{Parameter		& Value}	
\startdata
Right ascension (J2000) & $\rm 17^{\rm h} 56^{\rm m}46\fs6332(2)$\\
Declination (J2000) 	& $\rm -22^{\circ} 51'   59\farcs 4(2)$	\\

Galactic longitude (deg) & 6.50	\\
Galactic latitude (deg)  & +0.95	\\

Ecliptic longitude (deg) & +269.31 	\\
Ecliptic latitude (deg)  & 0.57	\\
\\
Period, $P$ (ms)	& 28.46158845494(2)	\\
Period derivative $\dot{P}$ (x $10^{-18}$) 
			& 1.0171(2)		\\

Epoch (MJD)		& 52086	\\
Dispersion Measure, DM (pc cm$^{-3}$) 
			& 121.18(2)	\\
\\
Orbital Period, $P_b$ (days) 
			& 0.319633898(2) \\
Eccentricity, $e$	& 0.180567(2)		\\
T$_0$ (MJD) 		& 52812.919653(1)	\\
Longitude of periastron, $\omega$ (deg)
			& 322.571(4)		\\
Projected semimajor axis, $a_1$sin$\,i$ (lt-s) 	
			& 2.7564(2)	\\
Advance of periastron, $\dot \omega$ (deg yr$^{-1}$)
			& 2.585(2)	\\
Gravitational redshift, $\gamma$ (ms)
			& 1.3(3)	\\
\\
Number of TOAs		& 1362			\\
Timing data span (MJD)	& 50996 -- 53176	\\
RMS timing residual ($\mu s$)	
			& 42			\\
\\
Flux density at 1400MHz (mJy) 
			& 0.6(1)		\\
Width of pulse at 50\%, $W_{50}$ (ms)		
			& 0.78			\\
Width of pulse at 10\%, $W_{10}$ (ms)		
			& 1.6			\\
\\
Characteristic Age, $\tau_c$ (Myr) 
			& 443			\\
Surface magnetic field, $B$ (Gauss) 
			& $5.4 \times 10^9$	\\
Total system mass, $m_1+m_2$ ($M_{\odot}$)
			& 2.574(3)		\\
Pulsar mass$^\dag$, $m_1$
			& $1.40^{+0.02}_{-0.03}$	\\
Companion mass$^\dag$, $m_2$
			& $1.18^{+0.03}_{-0.02}$	\\
Time to coalesence (Gyr)
			& 1.69		\\
Distance (kpc) - NE2001	& 2.5		\\
$|z|$ (kpc)		& 0.04		\\
\enddata
\label{t:parameters} 
\tablecomments{Values in parenthesis are twice the nominal
\textsc{TEMPO} uncertainties in the least significant digits quoted,
obtained after scaling time of arrival (TOA) uncertainties to ensure
$\chi^2=1$. Distance estimated from the `NE2001' Galactic electron
density model (Cordes \& Lazio 2002)\nocite{cl02a}. $^\dag$Pulsar and
companion masses (1-$\sigma$ errors) are derived from the DDGR model and
are highly correlated.}
\end{deluxetable} 

\clearpage

\begin{deluxetable}{lrrrrrrl}
\tablewidth{0pt}
\tablecaption{Binary systems containing radio pulsars which coalesce
in less than $10^{10}$~yr.}
\tablehead{
PSR	& $P$	& $P_b$	& $e$	& Total Mass	&$\tau_{\rm c}$&
$\tau_{\rm GW}$ & Reference\\
	& (ms)	& (hr)	& 	& $M_{\odot}$	&(Myr)	 & (Myr)	   } 

\nocite{bdp+03} \nocite{lbk+04} \nocite{wol90a} \nocite{ht75a}
\nocite{agk+90} \nocite{klm+00a}
\startdata

J0737$-$3039A	& 22.70	& 2.45	& 0.088	& 2.58 	& 210	& 87	& 
Burgay et al. (2003)\\
J0737$-$3039B	& 2773	& 2.45	& 0.088	& 2.58	& 50	& 87	& 
Lyne et al. (2004)	\\
B1534+12	& 37.90	& 10.10	& 0.274	& 2.75 	& 248	& 2690	& 
Wolszczan (1990)	\\
J1756$-$2251	& 28.46	& 7.67	& 0.181	& 2.57	& 444	& 1690	& 
This Letter	\\
B1913+16	& 59.03	& 7.75	& 0.617	& 2.83	& 108	& 310	& 
Hulse \& Taylor (1975)\\
B2127+11C	& 30.53	& 8.04	& 0.681	& 2.71	& 969	& 220	& 
Anderson et al. (1990	\\
J1141$-$6545$^{\dag}$ & 393.90  & 4.74 & 0.172	& 2.30	& 1.4 & 590 & 
Kaspi et al. (2000)	\\ 

\enddata \tablecomments{One neutron star--white dwarf$^{\dag}$ and 5
DNS systems. PSR B2127+11C is in a globular cluster implying a
different formation history to the Galactic DNS systems. Here,
$\tau_{\rm c}$ is the pulsars' characteristic age and $\tau_{\rm GW}$
is the time remaining to coalesce due to emission of gravitational
radiation. The total coalescence time is $\tau_{\rm c}+\tau_{\rm
GW}$.}
\label{t:coalescing} 
\end{deluxetable} 

\clearpage

\clearpage

\clearpage

\end{document}